\documentclass[conference, a4paper]{IEEEtran}
\IEEEoverridecommandlockouts
\usepackage{adjustbox}
\usepackage{cite}
\usepackage[utf8]{inputenc} 
\usepackage[T1]{fontenc}
\usepackage[english, turkish]{babel}
\usepackage{amsmath,amssymb,amsfonts}
\usepackage{amsthm}
\usepackage[ruled,vlined,linesnumbered,noresetcount]{algorithm2e}
\usepackage{graphicx}
\usepackage{textcomp}
\usepackage{xcolor}
\usepackage{xspace}
\usepackage{booktabs}

\usepackage{optidef}
\usepackage{algpseudocode}
\usepackage{lipsum}
\def\BibTeX{{\rm B\kern-.05em{\sc i\kern-.025em b}\kern-.08em
    T\kern-.1667em\lower.7ex\hbox{E}\kern-.125emX}}

\begin{document}

\title{Multi-Tier Non-Terrestrial Networking for Disaster Communications: A Layered Clustering Approach \\ 
Afet Haberleşmesi İçin Çok Katmanlı Karasal Olmayan Ağlanma: Katmanlı Kümeleme Yaklaşımı}


\author{\IEEEauthorblockN{Metin Ozturk$^{1,2}$, Berk Çiloğlu$^{1,2}$, Görkem Berkay Koç$^{1,2}$, Halim Yanikomeroglu$^2$}
\IEEEauthorblockA{$^1$Electrical and Electronics Engineering, Ankara Yıldırım Beyazıt University, Ankara, Türkiye\\
\IEEEauthorblockA{$^2$Non-Terrestrial Networks (NTN) Lab, Systems and Computer Engineering, Carleton University, Ottawa, Canada} \thanks{This research has been sponsored in part by The Scientific and Technological Research Council of Türkiye (TUBITAK).}}}
\maketitle

\selectlanguage{english}
\begin{abstract}
It is crucial to deploy temporary non-terrestrial networks (NTN) in disaster situations where terrestrial networks are no longer operable. 
Deploying uncrewed aerial vehicle base stations (UAV-BSs) can provide a radio access network (RAN); however, the backhaul link may also be damaged and unserviceable in such disaster conditions. 
In this regard, high-altitude platform stations (HAPS) spark attention as they can be deployed as super macro base stations (SMBS) and data centers. 
Therefore, in this study, we investigate a three-layer heterogeneous network with different topologies to prolong the lifespan of the temporary network by using UAV-BSs for RAN services and HAPS-SMBS as a backhaul.
Furthermore, a two-layer clustering algorithm is proposed to handle the UAV-BS ad-hoc networking effectively.
\end{abstract}
\begin{IEEEkeywords}
\textit{backhaul, disaster, HAPS, resilience, UAV}
\end{IEEEkeywords}

\selectlanguage{turkish}

\begin{ozet}
Karasal ağların artık çalışmayı durdurduğu afet durumlarında, geçici bir karasal olmayan ağların konuşlandırılması önemlidir. 
İnsansız hava aracı baz istasyonlarının (İHA-Bİ'ler) konuşlandırılması bir radyo erişim ağı (RAN) sağlayabilir ancak ana taşıyıcı bağlantısı da bu tür afet koşullarında hasar görmüş olabilir ve hizmet veremeyebilir. 
Bu bağlamda, yüksek-irtifa platform istasyonu (HAPS), süper makro baz istasyonu (SMBS) ve veri merkezi olarak konuşlandırılabileceği için dikkatleri üstüne çeker. 
Bu nedenle, bu çalışmada, RAN hizmetleri için İHA-Bİ ve ana taşıyıcı olarak HAPS-SMBS kullanarak geçici ağın ömrünü uzatmak için farklı topolojilere sahip üç katmanlı heterojen bir ağı araştırıyoruz. 
Ayrıca, afet senaryolarında, İHA-Bİ'lerin ad-hoc ağ haberleşmesini etkin bir şekilde ele almak için iki katmanlı bir kümeleme algoritması önerilmiştir.
\end{ozet}
\begin{IEEEanahtar}
afet, ana taşıyıcı, esneklik, HAPS, İHA
\end{IEEEanahtar}

\selectlanguage{english}

\IEEEpeerreviewmaketitle

\IEEEpubidadjcol

\section{Introduction}
In times of natural disasters and humanitarian crises, ensuring effective communication channels is crucial for coordinating search and rescue~(SAR) efforts, disseminating vital information, and providing essential services to people in affected regions. 
Since traditional communication infrastructures are often vulnerable to damage or complete collapse during such events, the use of uncrewed aerial vehicles~(UAVs), as non-terrestrial networks~(NTN) element, emerges as a promising solution to meet communication needs~\cite{UAV_disaster}.
This is achieved by deploying base stations~(BSs) on UAVs, which are referred to as UAV-BSs.
UAVs can be regarded as an ideal choice for SAR missions, disaster management, and emergency response operations due to their speed, flexibility, and mobility~\cite{UAV_disaster}, and they can be easily deployed and positioned anywhere, even in hard-to-reach areas, providing communication connectivity. 

Nonetheless, on the flip side of the coin, UAVs need proper backhauling strategy in the event of aforementioned disasters, as the terrestrial network components may be damaged and/or fully collapsed.
At this point, the use of high altitude platform stations (HAPS), another NTN element, can be considered as a viable solution, given that they provide high capacity, ubiquitous connectivity, and a powerful data center to the network due to the super macro BS~(SMBS)\footnote{International Telecommunication Union (ITU) makes reference HAPS-SMBS as high-altitude IMT BS (HIBS)~\cite{itu_vision_june_23}.} it contains~\cite{haps-smbs}.
HAPS is located in the stratosphere, at an altitude of approximately 20 km above the Earth's surface. 
This strategic positioning allows HAPS to provide high-quality communication services to a large geographical area without repositioning~\cite{main_survey}. 

Although extensive research has been conducted on the use of terrestrial BSs for backhauling UAVs during disaster scenarios, studies on the use of HAPS-SMBS for backhauling purposes in such cases are still in their infancy. 
The study in~\cite{Backhaul_Cons} analyzed the network coverage based on high-altitude platforms-assisted backhaul and low-altitude platforms direct backhaul strategies in post-disaster areas. 
The authors in~\cite{Omid} studied on the UAV-BSs through HAPS-assisted backhauling within a cell-free scheme. 
A framework for planning and evaluating aerial platform-based wireless backhaul networks was proposed in~\cite{wireless_bachaul}, wherein a number of variables such as altitude, platform type, deployment strategy, energy management, and security were considered.
The study in~\cite{IoT_Collecting} focused on collecting internet of things~(IoT) data in a disaster scenario using UAVs and HAPS-SMBS.

This current study is one of the leading works to mention that HAPS-SMBS can create a backhaul link for UAVs in the context of disaster scenario.
In particular, a three-layer network is proposed, wherein the user equipments~(UEs) are on the first layer (i.e., ground), UAV-BSs are on the second layer (i.e., 100-200 meters above the ground) providing radio access network~(RAN) services, and HAPS-SMBS is on the third layer (i.e., around 20 km from the ground) providing backhaul services to the UAV-BSs.
In addition, a two-layer clustering algorithm is proposed, such that UEs, who gather in hot-spots, form clusters on the first layer, and the UAV-BSs on the second layer are positioned according to the results of this clustering. 
Then, a further clustering operation is also performed on the UAV-BSs for the backhaul link.
More specifically, the cluster head UAV, referred to as H-UAV, collects data from other UAVs in the form of \textit{ad-hoc} networking and transmit to HAPS-SMBS.
In other words, only H-UAV---positioned at a pre-designated location that has a grid power connection as well as uninterrupted power supply---has a backhaul link to prolong the network lifespan, because the energy consumption of non-head UAVs~(referred to as NH-UAVs) for the backhaul connection is eliminated in our proposed topology.

As we consider a  disaster scenario where providing connectivity is of paramount importance, we do not target to minimize the overall energy consumption of the network; instead, we aim to maximize the network lifespan to provide communication services for a longer time by minimizing the energy consumption of network components that rely solely on battery energy sources.
  
\section{System Model}
\begin{figure*}[hbt!]
    \centering
     \shorthandoff{=}
     \includegraphics[width=.6\linewidth,trim={1cm 0cm 1cm 1cm},clip]{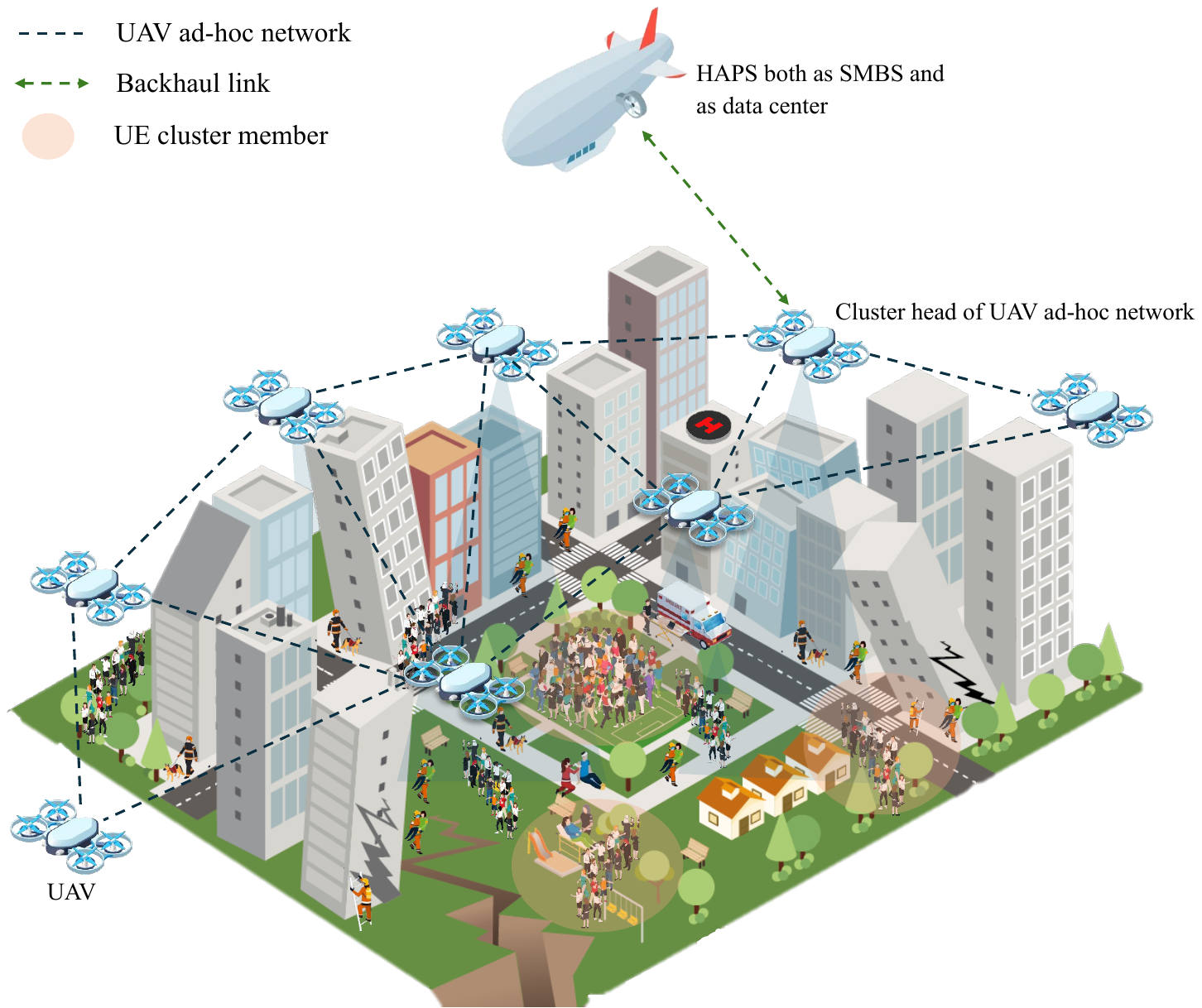}
    \shorthandoff{=}
    \caption{A city model is formed during a disaster, infrastructures have collapsed and communication is provided by UAVs, HAPS is backhauling the UAVs, and UEs are clustered in places such as gathering areas.}
    \label{fig:networkmodel}
\end{figure*}

\subsection{Network Model}
A three-tier vertical heterogeneous network~(VHetNet) including $n \in \mathbb{N}$ UAV-BSs, with $j=\{1, 2, ..., n\}$ storing the indices of UAVs, and a single HAPS-SMBS is considered as a network model. 
The environment area is 3 km $\times$ 3 km.
There are also $e \in \mathbb{N}$ users/UEs, with $i=\{1, 2, ..., e\}$ storing the indices for the users/UEs, distributed around the environment by forming clusters.
HAPS is located 20 km above the center of the environment, while the UAVs are distributed around the environment with an altitude of 150 m.
There are also landing spots, powered by grid and generators, in some parts of the city.
The system model is illustrated in Fig.~\ref{fig:networkmodel}.

\subsection{Energy Consumption Model}
We are mainly interested in the energy consumption of the UAVs in the network as they are mostly battery-powered and can create a bottleneck in the communication service time.
The total energy consumption of a UAV, $E_\text{T}$, is usually modeled by the accumulation of the energy consumption values due to communication, $E_\text{C}$, and flight, such as hovering and mobility~\cite{UAV_energy}.
In our scenario, we aim to investigate the dynamics of the total energy consumption of the UAV network; since the communication energy consumption is the predominant factor, and the flight energy consumption remains consistent across all comparisons, we solely focus on the communication~energy.

The variable $E_\text{C}$ is a function of transmit power and load of a BS, as stated in~\cite{UAV_energy}, that is, $E_\text{C} = f(P_\text{tx},\lambda)$, where $P_\text{tx}$ is the transmit power and $\lambda$ is the load of the BS, while $f: \mathbb{R}^+ \rightarrow \mathbb{R}^+$ is a function.
A power control strategy is adopted in this work, such that the received power, $P_\text{rx}$, is fixed to the receiver sensitivity of the UEs and then $P_\text{tx}$ is determined accordingly (i.e., by considering the path loss, $L$, and $P_\text{rx}$). 
In other words, $P_\text{tx}$ is a function of both $P_\text{rx}$ and $L$; i.e., $P_\text{tx}=f(P_\text{rx},L)$.
Moreover, $L$ is a usually a distance-dependent variable, such that $L=f(d)$, thereby $P_\text{tx}$ can also said to be distance-dependent, that is, $P_\text{tx} = f(d)$.
Therefore, at the abstract level of energy consumption calculation, $E_\text{C}$ is a function of $d$ and $\lambda$, that is, $E_\text{C} = f(d,\lambda)$.
Thus, we define a metric called \textit{energy score}, $E_\text{s}$, to represent the abstract level energy consumption of a UAV-BS as
\begin{equation} \label{eq:energy}
    E_\text{s} = \lambda d, 
\end{equation} 
where $d$ is the distance between the UAV-BS and a receiver.

\section{Multi-Tier Non-Terrestrial Networking for Disaster Communications}
We propose a two-layer clustering algorithm for positioning the UAV-BSs and the backhauling between the UAV-BSs and HAPS-SMBS.
The first-layer clustering performed for positioning the UAV-BSs.
For that, UEs are arranged into groups, concentrated in hotspots, followed by employing a $k$-means algorithm to determine the centroids of these UE groups. 
In particular, the two-dimensional coordinates of the UEs are fed into the $k$-means algorithm to find the centroid positions, which are then used as the two-dimensional positioning of UAV-BSs (the altitude of the UAV-BSs is fixed to 150 m).

The second-layer clustering is then applied to find the communication topology for backhauling.
Unlike the first-layer clustering, this time the two-dimensional coordinates of the UAV-BSs (obtained in the first-layer clustering) are fed into another $k$-means algorithm to compute the centroid positions.
After that, the UAV-BS that has the closest proximity to the centroid is assigned as H-UAV, and the rest of the UAV-BSs serve as NH-UAVs.
The UAV-BSs then form an ad-hoc network in such a way that the NH-UAVs are transferring their data to the H-UAV, which has the capability of connecting to HAPS-SMBS via a backhauling link.
As mentioned earlier, some landing spots, with grid and generator power supplies as well as a clear line-of-sight with the majority of the area that they are planned to serve, are deployed around the city.
Therefore, the H-UAV is positioned at the nearest landing spot for continuous energy supply to prevent its battery from depleting fast as it is responsible to transmit all the data of the affected region.
The effects of the distance between the landing spots and optimal position of UAV-BSs on the energy efficiency are deeply investigated in~\cite{landing_UAV}, which can be used as a complementary source to this present work.

Using the logic in~\eqref{eq:energy}, the energy score of the proposed method can be computed as 
\begin{equation}\label{eq:energy_proposed}
    E_\text{p} = \sum_{j=1}^{n-1} (\lambda_j \overline{d_{j,e}}+\lambda_k d_{j,\text{H}}),
\end{equation}
where $\overline{d_{j,e}}$ is the average distance between the UAV-BS $j$ and all the UEs, and $\lambda_j$ is total load of UAV-BS $j$ while $d_{j,\text{H}}$ is the distance between UAV-BS $j$ and the H-UAV.
Since the H-UAV has a continuous power supply on the landing spot and this study focuses on prolonging the service time of the UAVs, the energy score of the H-UAV is excluded in~\eqref{eq:energy_proposed}.

\section{Performance Evaluation}
\subsection{Benchmarking}
In addition to the above-mentioned proposed topology, referred to as \textit{double-layer clustering approach with ad-hoc networking (DLC-AHN)} hereafter, two different benchmark topologies are also considered.
The first benchmark is that UAV-BSs are positioned in the same way as DLC-AHN (i.e., with the $k$-means clustering); however, this time each UAV-BS connects with HAPS-SMBS for backhauling purposes. 
This benchmark is referred to as \textit{single-layer clustering~(SLC)} approach.
In the second benchmark, referred to as \textit{circular UAV positioning (CUP)}, the UAV-BSs are distributed in a circular manner around the environment, and each UAV-BS establishes a backhaul connection with HAPS-SMBS.

For both benchmark methods (i.e., $E_\text{SLC}$ for SLC and $E_\text{CUP}$ for CUP) the total energy score is obtained as
\begin{equation} \label{eq:energy_23}
   E_b = \sum_{j=1}^{n-1} \lambda_j ( \overline{d_{j,e}}+ d_{j,\text{S}}),
\end{equation}
where $b \in \{\text{SLC},\text{CUP}\}$ and $d_{j,\text{S}}$ is the distance between UAV-BS $j$ and HAPS-SMBS. 
The network is evaluated for varying UE densities (users/m$^2$ ratios), denoted by $\delta$. 

\subsection{Results and Discussions}
\begin{figure}[b!]
    \centering
     \shorthandoff{=}
    \includegraphics[width=.75\linewidth]{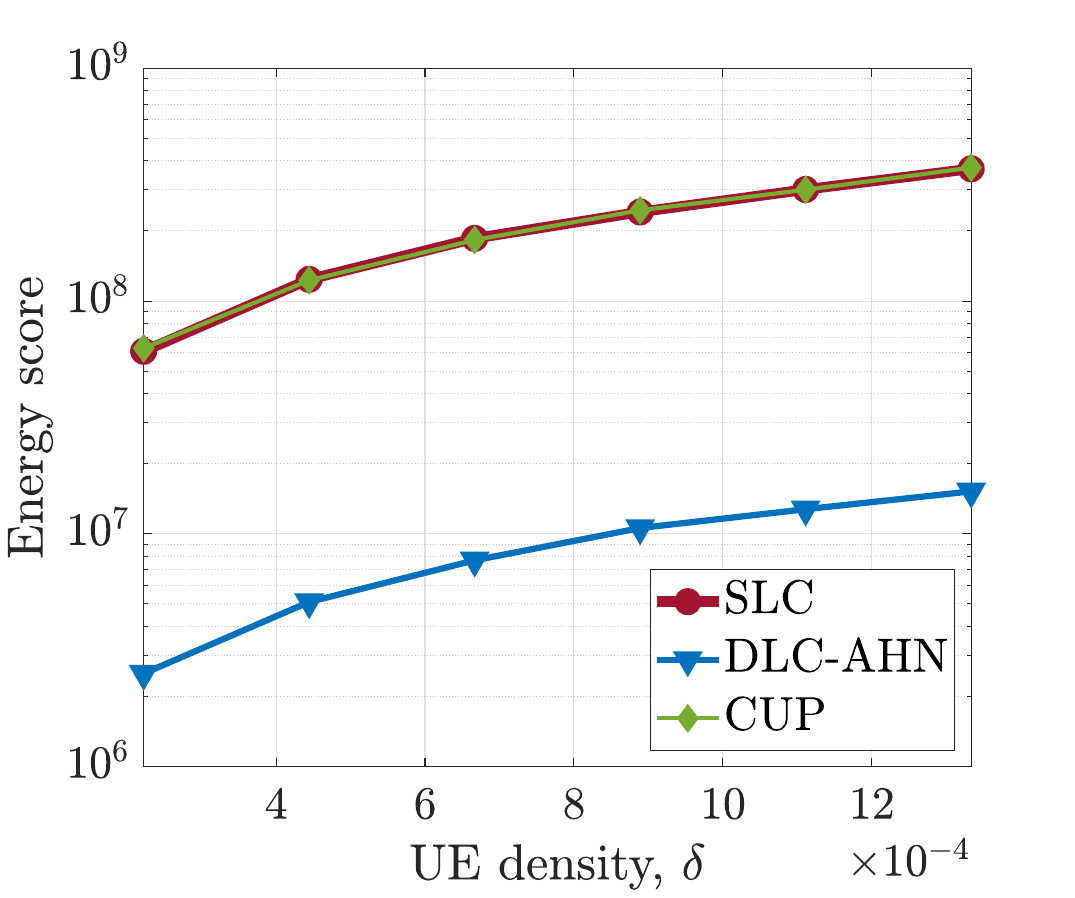}
    \shorthandoff{=}
    \caption{Energy score with respect to the user density.}
    \label{fig:energy}
\end{figure}
Fig.~\ref{fig:energy} demonstrates the energy scores for all scenarios across various $\delta$ values.
As expected, the energy score soars with $\delta$ for each method as $\lambda$ rises with increasing $\delta$, which in turn boosts the total energy score as given in~\eqref{eq:energy_proposed} and \eqref{eq:energy_23}.
While SLC and CUP show similar behaviors, DLC-AHN is separated from them with a considerably lower energy score. 
The reason for this is that the NH-UAVs do not have backhauling link to HAPS-SMB and they only transmit their data to H-UAV, which is in a much closer proximity to them compared to HAPS-SMBS.
For SLC and CUP, on the other hand, all the UAV-BSs need to transmit their data to HAPS-SMBS, and considering \eqref{eq:energy_23}, their energy scores soar significantly.
SLC and CUP appear almost on top of each other in Fig.~\ref{fig:energy}, but there is a difference of 0.41\% between them in terms of the total energy score.
In SLC, since UAV-BSs are positioned at the centroid point determined as a result of the clustering, the distance is minimized and they consume less energy than CUP.

\begin{figure}[t!]
    \centering
     \shorthandoff{=}
    \includegraphics[width=.74\linewidth]{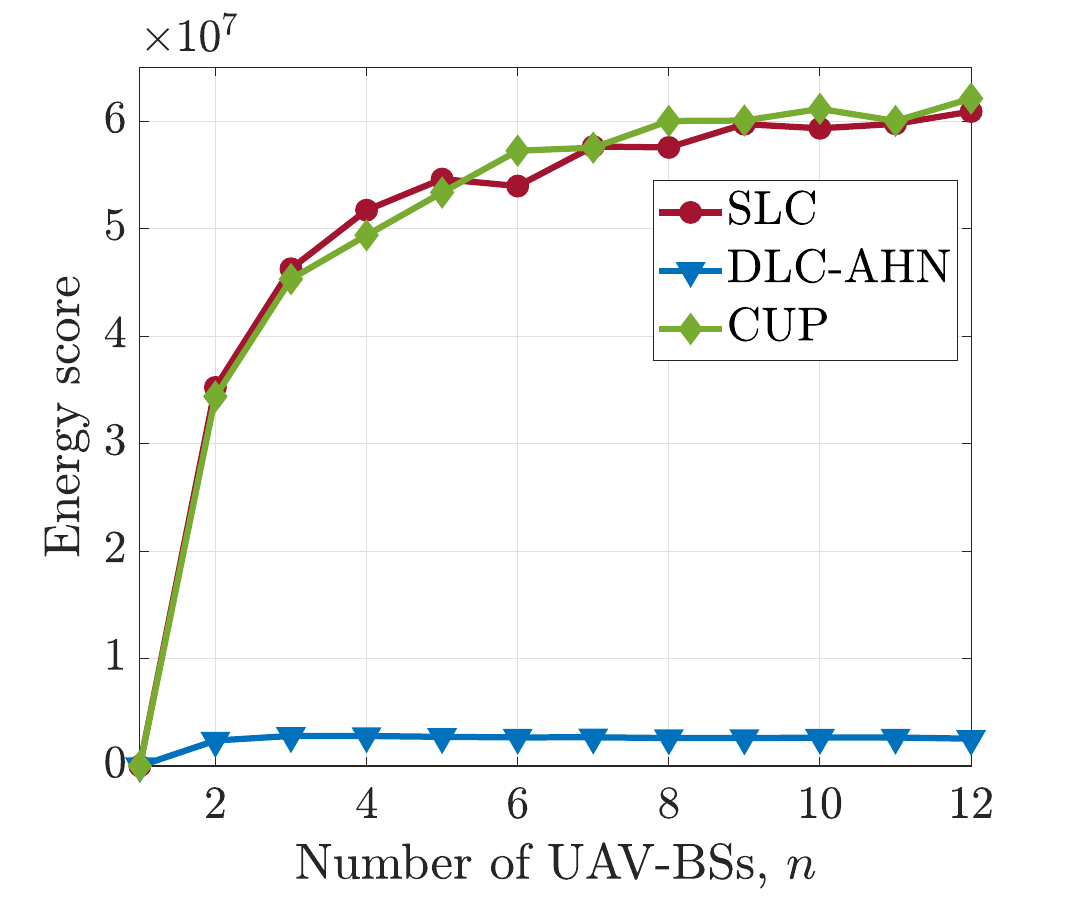}
    \shorthandoff{=}
    \caption{Energy score for varying number of $n$ in the first-layer clustering.}
    \label{fig:energy2}
\end{figure}

Fig.~\ref{fig:energy2} displays the energy scores with respect to the number of UAV-BSs.
The generic trend that is common to all the methods is that the energy score increases with the number of UAV-BSs (or the number of clusters in the first-layer clustering).
Note that the energy score of DLC-AHN is increasing despite appearing constant due to visualization scaling.
Considering \eqref{eq:energy_proposed} and \eqref{eq:energy_23}, this outcome is anticipated because the energy score scales with the number of UAV-BSs.
However, DLC-AHN not only results in lower energy scores but also keeps the level of rise with the increasing number of UAV-BSs under control compared to SLC and CUP, which demonstrate a dramatic surge in the energy score as a response to the increasing number of UAV-BSs. 
This is attributed to the distance variable, $d$, in \eqref{eq:energy_proposed}, since when the number of UAV-BSs increases the distance between them decreases, preventing the overall sum in \eqref{eq:energy_proposed} going up sharply as SLC and CUP. 

\begin{figure}[h!]
    \centering
     \shorthandoff{=}
    \includegraphics[width=.74\linewidth]{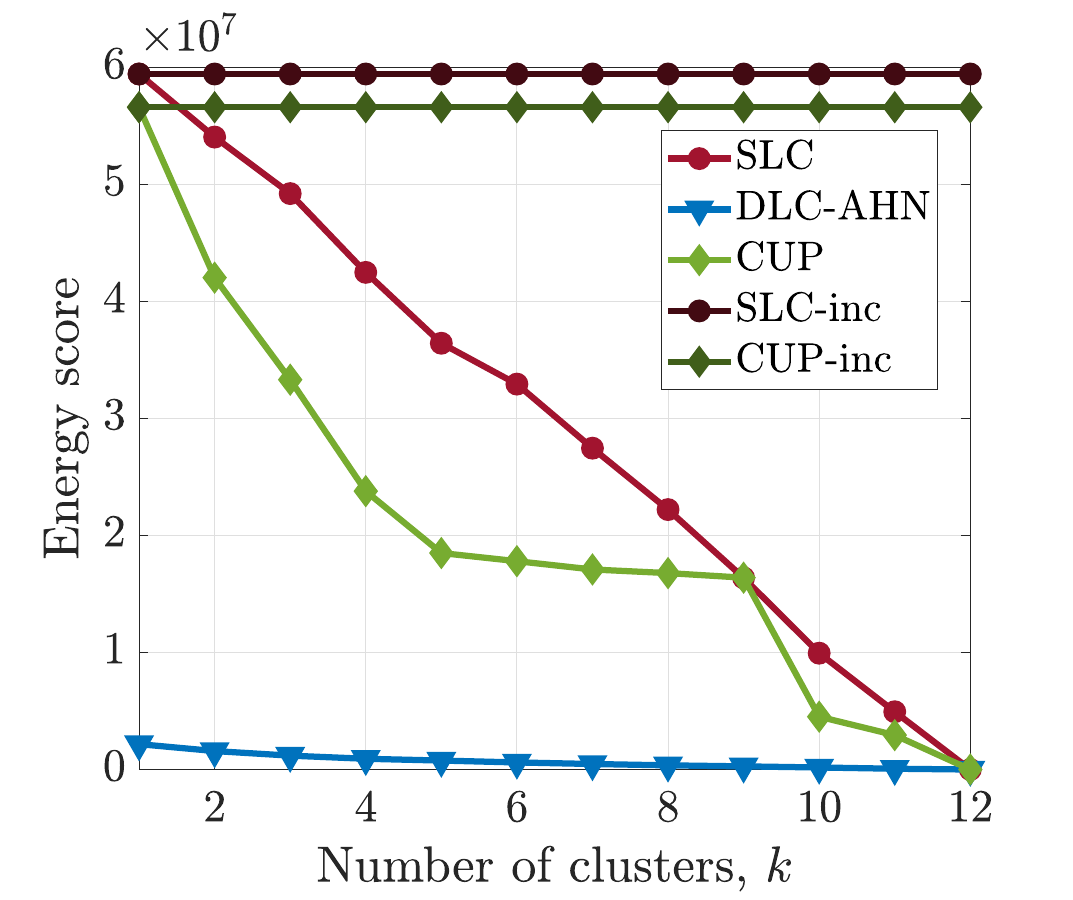}
    \shorthandoff{=}
    \caption{Energy score for different $k$ values in the second-layer clustering.}
    \label{fig:energy3}
\end{figure}

Fig.~\ref{fig:energy3} illustrates the energy score metric with respect to $k$ values in the second-layer clustering. 
To make a fair comparison between the benchmarks, the H-UAVs selected for DLC-AHN are omitted in the energy score calculations for all the methods.
\textit{SLC-inc} and \textit{CUP-inc} in Fig.~\ref{fig:energy3} represent the case where omitting the H-UAVs in the energy calculations is not practiced.
As seen, \textit{SLC-inc} and \textit{CUP-inc} results are flat as the second-layer clustering is not applied to them.
Considering all three methods, it is clear to see that there are decreases in the energy score.
The reason behind this is that the number of H-UAVs increases with $k$, and since the energy scores of H-UAVs are not included, the overall energy consumption approaches to zero.
In other words, when the number of cluster increases, the UAV-BSs without grid and/or generator power supply decreases, thereby the issue of service time disappears.

\section{Conclusion}
In this study, we analyzed the concept of using HAPS-SMBS as a backhaul link for the network of UAV-BSs in disaster circumstances.
We proposed a two-layer clustering algorithm aimed at enhancing network survival time and effectively managing communication under unfavorable network conditions.
The first-layer clustering is utilized for positioning the UAV-BSs, which are employed for RAN services, while the second-layer clustering is used for backhauling purposes.
The results obtained confirm that our three-layer network with a two-layer clustering approach has the potential to extend communication service time during disaster scenarios, providing rapid recovery and network resilience.

\bibliographystyle{IEEEtran}
\bibliography{output}
\end{document}